# Coupled mechano-electrokinetic Burridge-Knopoff model of fault sliding events and transient geoelectric signals

Authors: Hong-Jia Chen[1], Chien-Chih Chen[1,4], Guy Ouillon[2], Didier Sornette[3]

Affiliations:

[1] Department of Earth Sciences, National Central University, Zhongli, 32001, Taiwan.

[2] Lithophyse, Nice, France.

[3] Department of Management, Technology and Economics, ETH Zürich, Zürich, Switzerland.

[4] Earthquake-Disaster & Risk Evaluation and Management Center, National Central University, Zhongli, 32001, Taiwan.

Corresponding author: Hong-Jia Chen (redhouse6341@g.ncu.edu.tw)

# Abstract

We introduce the first fully self-consistent model combining the seismic micro-ruptures occurring within a generalized Burridge-Knopoff spring-block model with the nucleation and propagation of electric charge pulses within a coupled mechano-electrokinetic system. This model provides a general theoretical framework for modeling and analyzing geoelectric precursors to earthquakes. In particular, it can reproduce the unipolar pulses that have often been reported before large seismic events, as well as various observed anomalies in the statistical moments of the ambient electric fields and the power-law exponent transition of the power spectra of electric fields.

# 1. Introduction

Geoelectric signals and related transient anomalies have been proposed for decades as potential precursors to large earthquakes, but the seismological community is still skeptical due to the lack of an established clear theoretical relationship between crustal rock mechanics and electrics within the crust. Such physical mechanisms are still debated, and a few plausible models have been proposed. Those models feature ingredients such as electrokinetic effect (Ishido & Mizutani, 1981; Mizutani et al., 1976; Yamada et al., 1989), piezoelectric effect (Ikeya et al., 2000; Nitsan, 1977; Sasaoka et al., 1998; Sornette & Sornette, 1990), pressure-stimulated current (Stavrakas et al., 2004; Vallianatos & Triantis, 2008; Varotsos & Alexopoulos, 1984), or stress-activated peroxy defects (Freund, 2003, 2007; Freund et al., 2009; Freund & Pilorz, 2012; Freund & Sornette, 2007). The later mechanism has been the most developed, based both on detailed experimental tests and observational studies. It will thus be the center of attention of the present work.

In contrast, the mechanics of individual earthquakes is believed to be much better understood, not only because of a longer history of observations and recordings of waveforms, but also due to the huge efforts the community put in computer modeling. In an effort to capture the essence of the self-organization of earthquake sequences, Burridge and Knopoff (1967) proposed a simple spring-block model to simulate the dynamics of a fault. This model has triggered many other works and publications (Abaimov et al., 2007; Andersen et al., 1997; Brown et al., 1991; Cao & Aki, 1984; Carlson, 1991; Carlson et al., 1991, 1994; Carlson & Langer, 1989a; Cartwright et al., 1997; Chen et al., 2012; Chen & Wang, 2010; Erickson et al., 2008, 2011; Hasumi et al., 2010; Mitsui & Cocco, 2010; Nussbaum & Ruina, 1987; Wang, 2012; Wang, 2008; Wang, 2009; Wang & Hwang, 2001; Yoshida & Kato, 2003). In those studies, the spring-block models are used to exhibit some specific phenomena, such as the power-law distribution of event sizes, slip complexity (Wang & Hwang, 2001), the aftershocks caused by material decoupling (Chen et al., 2012), propagating slip pulses (Chen & Wang, 2010), the tricritical behavior of rupture (Andersen et al., 1997), and so on.

If the main features of earthquake rupture are reasonably encapsulated in the simple mechanical models described above, the peroxy-defects theory proposed by Freund and his collaborators is less advanced and only provides a thread to build up models of the behavior of electric charges within the crust. In Freund's theory, peroxy bonds in crustal rocks are considered to be perturbed and broken during ever-increasing tectonic stresses prior to any major seismic activity, so that electric charges (electrons and positive holes) in the rock minerals would be activated. The positive holes are able to flow out of the stressed rock volume, into and through surrounding unstressed or less stressed rock, forming observable electric currents.

Here, we develop a new macroscopic conceptual model that couples the mechanical and electrical behaviors of a discrete set of blocks connected by elastic springs, using RLC circuit elements to describe the flow of electric charges, as well as a novel source term to embody the mechanical-electrical coupling. We thus present the first integrated model of mechanical earthquake ruptures coupled with electric charge production and transport, in the goal of studying how the observed electric signals could reveal the earthquake spatiotemporal organization, and especially the upcoming of the largest events. Our work also provides both theoretical and phenomenological evidences for correlations between the anomalies of electric fields and the occurrences of seismic slips at different spatial scales. The emerging signals may then be used in future works to search empirically for similar anomalies in time series of naturally occurring electric fields, and use them in order to predict large earthquakes (see Bleier et al., 2009; Chen et al., 2017; Chen & Chen, 2016; Eftaxias et al., 2001, 2003; Potirakis et al., 2013; Scoville et al., 2015; Telesca et al., 2004, 2014; Varotsos, 2005; see also our companion paper in this volume).

The organization of this article is as follows. The next section summarizes the main building elements of the standard spring-block system used classically to model earthquake events as fault slidings and their spatio-temporal organization. Section 3 shows how to represent the creation of free charges by the breaking of peroxy bonds within a single block picture. Section 4 introduces the multi-blocks problem, its general equations and solutions. Section 5 concludes by presenting a discussion linking our findings to the existing literature.

## 2. Spring-block system

To begin with, we introduce a one-dimensional spring-block system. We consider a linear chain of $N$ blocks of identical mass $m$ pulled over an interface at a velocity $v_L$ by a loading plate as shown in Fig. 1b. Each block is connected to the loading plate by a spring with stiffness $K_L$, while adjacent blocks are connected to each other by a spring with stiffness $K_C$. The definition of the boundary conditions is a subtle problem. The boundary conditions should refer to the real geometry, and have been fully discussed in the two-dimensional case (Christensen & Olami, 1992b, 1992a). For the one-dimensional case, it has been suggested that spring-block models with different boundary conditions yield similar results (Carlson & Langer, 1989b). In our study, geometrical boundary conditions are assumed to be periodic so that the $N^{th}$ block is linked with the $1^{st}$ one.

The blocks slide over a perfectly flat frictional interface. The static stability condition for each block is given by:

$$K_L x_i + K_C(2x_i - x_{i-1} - x_{i+1}) < F_{Si}, \quad i = 1 \text{ to } N, \tag{1}$$

where $F_{Si}$ is the static frictional threshold force of the $i^{th}$ block, and $x_i$ is the position of the $i^{th}$ block relative to the loading plate. During strain accumulation due to the loading by the plate motion, all blocks are motionless relative to the interface and witness the same increase of their coordinates relative to the loading plate:

$$\frac{dx_i}{dt} = v_L, \quad i = 1 \text{ to } N, \tag{2}$$

When the resulting force of the springs connected to the $i^{th}$ block exceeds the static threshold $F_{Si}$, the block begins to slide. The dynamic slip of the $i^{th}$ block, including inertia effects, is now given by

$$m\frac{d^2 x_i}{dt^2} + K_L x_i + K_C(2x_i - x_{i-1} - x_{i+1}) = F_{Di}, \quad i = 1 \text{ to } N, \tag{3}$$

where $F_{Di} < F_{Si}$ is the dynamic frictional force acting on the $i^{th}$ block. The sliding of one block can trigger the instability of the other blocks, thus forming a multi-blocks event. When the velocity of a block is zero, it sticks to the interface with zero velocity if the static friction criterion Eq. (1) is satisfied; if not satisfied, the block continues to slip according to Eq. (3).

In order to scale the above-mentioned equations, we introduce the following dimensionless variables and parameters:

$$T_f = t\sqrt{\frac{K_L}{m}}, \; T_s = \frac{tK_L v_L}{F_S^{ref}}, \; X_i = \frac{K_L x_i}{F_S^{ref}}, \; \phi = \frac{F_{Si}}{F_{Di}}, \; s = \frac{K_C}{K_L}, \; \mu_i = \frac{F_{Si}}{F_S^{ref}}. \tag{4}$$

The $s$ is the stiffness ratio, representing the level of conservation of energy in the system. A larger stiffness ratio indicates a higher level of conservation or a lower level of dissipation of energy, while the probability of multi-blocks, larger-sized events increases with $s$ (Wang & Hwang, 2001). The ratio $\phi$ of static to dynamic frictional forces is assumed to be the same for all blocks, but $\mu_i$ varies from block to block with $F_S^{ref}$ being a reference value for the static frictional force (here, the minimum value of all the $F_{Si}$'s). Stress accumulation takes place during the 'slow time $T_s$' when all blocks are stable, and sliding of blocks occurs during the 'fast time $T_f$' during which the loading plate is assumed to be approximately immobile. In terms of these dimensionless variables and parameters, the static stability condition Eq. (1) becomes:

$$X_i + s(2X_i - X_{i-1} - X_{i+1}) = \tau_i < \mu_i, \quad i = 1 \text{ to } N, \qquad (5)$$

where $\tau_i$ stands for the stress acting on the $i^{th}$ block. The strain accumulation Eq. (2) becomes:

$$\frac{dX_i}{dT_s} = 1, \quad i = 1 \text{ to } N. \qquad (6)$$

The dynamic slip Eq. (3) becomes:

$$\frac{d^2 X_i}{dT_f^2} + X_i + s(2X_i - X_{i-1} - X_{i+1}) = \frac{\mu_i}{\phi}, \quad i = 1 \text{ to } N \qquad (7)$$

Finally, the total amount of slip within the spring-block system is defined as:

$$D_{SB}^{(t)} = \sum_{i=1}^{N} \left( X_i^{(t)} - X_i^{(t-1)} \right), \qquad (8)$$

where $t$ stands for the slipping time points. Table 1 lists the definitions and values of the spring-block parameters. In the numerical simulations, we specify the parameters $N$, $\phi$, $s$, and $\mu_i$. In this work, we set $N=128$, $s=30$, $\phi=1.5$, while the $\mu_i$'s are assigned to blocks using a uniform random distribution within the range $1<\mu_i<3.5$.

## 3. Electrokinetic system: single-block problem

### 3.1 Description and governing equations

Experiments on positive hole charge carriers in rocks (Freund, 2003, 2007; Freund et al., 2009; Freund & Pilorz, 2012) provide evidence that the production of electric charges (hence voltage, the equivalent of an electrical pressure) is proportional to the applied stress, due to the constant resistance of the compressed material (see Fig. 4 of Freund, 2007). Hence, we consider that the mechanical and electrical variables within the crust are coupled through a stress-induced voltage ($V_{in}$),

such that:

$$V_{in}(\tau) = \beta \cdot \tau, \tag{9}$$

where $\beta$ is a positive constant and $\tau$ is the stress. The unit of $\beta$ is mV/MPa, based on the results of Takeuchi et al. (2006, see Fig. 5 therein): Considering a granite sample, a stress variation of 50 MPa can produce a voltage change of approximately 40 mV. Note that, in the numerical simulations, we consider a dimensionless parameter $\beta$.

According to those experimental results, we assume that each block plays the role of a resistor with resistance $r$ and of a capacitor with capacitance $c$, as shown in Fig. 1a. The block resistance and capacitance would be influenced by petro-fabric, pore fabric, salinity of pore fluid, etc. (Nabawy, 2015). The block capacitor charges or discharges depending on the stress acting on the block. On the other hand, the block is embedded in the Earth's crust, i.e. is electrically grounded. The grounded current ($I$) passes through a grounded resistor with resistance $R$ and a grounded inductor with inductance $L$. The grounded resistance is an ambient resistance to the block, and the grounded inductance, related to the permeability of rock materials, is the ability to transform magnetic energies by flowing currents.

According to the above-mentioned scheme, the equations of the RLC-type circuit for $N=1$ (Fig. 1a) are derived as follows. First, Kirchhoff's voltage law in the block provides:

$$V_{in} - i_r r - \frac{q}{c} = 0. \tag{10}$$

Second, the current-charge relation in the block capacitor yields:

$$i_c = \frac{dq}{dt}. \tag{11}$$

Third, using Kirchhoff's current law between the block and the ground (at node A in Fig. 1a), we get:

$$i_r = I + i_c. \tag{12}$$

Finally, the equality of the voltage of the block capacitor and the grounded component (using nodes A and B in Fig. 1a) gives:

$$IR + \frac{dI}{dt}L = \frac{q}{c}. \qquad (13)$$

In Eqs. (10)-(13), $i_r$ stands for the current flowing away from the anode and passing through the block resistor ($r$), $q$ for the stored charges of the block capacitor ($c$), and $i_c$ for the current flowing towards the block capacitor ($c$). The unknown time-dependent variable vector is $\boldsymbol{G}=[q, i_c, i_r, I]$ with the initial condition $\boldsymbol{G}(t=0)=\vec{0}$.

Thus far, we have conceptualized a new model combining the mechanics of stick-slip in a spring-block system with the generation and propagation of electric charges within a coupled RLC circuit, which we refer to as the Chen-Ouillon-Sornette (COS) model hereafter. The mechanical component of the system is essentially a one-dimensional Burridge-Knopoff model, which is used to simulate stick-slip motions and earthquake ruptures. On the other hand, the electrokinetic component consists of a series of RLC-type circuits, while the peroxy-defects theory of Freund (2007) is used to motivate the description of the coupling between the stress acting on blocks with the amount of electric charges newly created. The model is sketched in Fig. 1a for a single-block case and in Fig. 1b for a multi-blocks case (see Section 4 for a more detailed description of the latter).

## 3.2 Analytical solutions

For Eqs. (10)-(13), we solve the problem analytically in the single-block case. In order to scale the four equations, we introduce the following dimensionless variables and parameters:

$$T = \frac{t}{c_{ref}R}, \quad \hat{r} = \frac{r}{R}, \quad \hat{c} = \frac{c}{c_{ref}}, \quad \hat{L} = \frac{L}{c_{ref}R^2}, \quad \widehat{V_{in}} = \frac{V_{in}}{i_{ref}R}, \quad \hat{q} = \frac{q}{i_{ref}c_{ref}R}, \quad \hat{i}_r =$$

$$\frac{i_r}{i_{ref}}, \quad \widehat{i_c} = \frac{i_c}{i_{ref}}, \quad \hat{I} = \frac{I}{i_{ref}}. \tag{14}$$

In Eqs. (14), $c_{ref}$ stands for a reference capacitance in SI unit of farad, and $i_{ref}$ for a reference current in SI unit of ampere. Therefore, the dimensionless electrokinetic equations become:

$$\widehat{V_{in}} - \hat{r}\widehat{i_r} - \frac{\hat{q}}{\hat{c}} = 0,$$

$$\widehat{i_c} = \frac{d\hat{q}}{dT},$$

$$\widehat{i_r} = \hat{I} + \widehat{i_c},$$

$$\hat{I} + \hat{L}\frac{d\hat{I}}{dT} = \frac{\hat{q}}{\hat{c}}. \tag{15}$$

In order to study the Green's function of such an electrokinetic system, we set $\widehat{V_{in}}(t)$ as a Dirac delta function $\delta(t)$, and use the Laplace transform rather than the Fourier transform because the electric behavior of the system is transient, not periodic. Taking the Laplace transform of Eqs. (15), they become, respectively:

$$1 - \hat{r}\tilde{i_r} - \frac{\tilde{q}}{\hat{c}} = 0,$$

$$\tilde{i_c} = s\tilde{q},$$

$$\tilde{i_r} = \tilde{I} + \tilde{i_c},$$

$$\tilde{I} + \hat{L}s\tilde{I} = \frac{\tilde{q}}{\hat{c}}, \tag{16}$$

where $\tilde{f}(s) = L[f(t)]$ is the notation for the Laplace transform. Note that the initial conditions are set to $\vec{G}(t=0)=\vec{0}$. After combination of these equations, we get:

$$\tilde{q}(s) = \frac{\hat{c} + \hat{c}\hat{L}s}{(1+\hat{r}) + (\hat{L}+\hat{r}\hat{c})s + (\hat{r}\hat{c}\hat{L})s^2}, \tag{17}$$

which finally yields:

$$\tilde{q}(s) = \frac{\left(s + \frac{1}{\hat{L}}\right)}{\hat{r}\left[s^2 + \left(\frac{1}{\hat{r}\hat{c}} + \frac{1}{\hat{L}}\right)s + \left(\frac{1}{\hat{r}\hat{c}\hat{L}} + \frac{1}{\hat{c}\hat{L}}\right)\right]}. \tag{18}$$

By defining $\zeta = \frac{1}{\hat{r}\hat{c}} + \frac{1}{\hat{L}}$, $\eta = \frac{1}{\hat{r}\hat{c}\hat{L}} + \frac{1}{\hat{c}\hat{L}}$, and $\Delta = \zeta^2 - 4\eta$, using inverse Laplace transform, we obtain three different cases for the Green function $q_{gf}(T)$ as follows:

**Case 1 ($\Delta$>0) - overdamping solution:**

$$q_{gf}^{(od)}(T) = \frac{e^{\frac{-\zeta+\sqrt{\Delta}}{2}T}\left(\frac{-\zeta+\sqrt{\Delta}}{2}+\frac{1}{\hat{L}}\right)+e^{\frac{-\zeta-\sqrt{\Delta}}{2}T}\left(-\frac{1}{\hat{L}}-\frac{-\zeta-\sqrt{\Delta}}{2}\right)}{\hat{r}\sqrt{\Delta}}, \tag{19}$$

where the characteristic decay time is $\tau_q = \frac{2}{\zeta-\sqrt{\Delta}}$.

**Case 2 ($\Delta$=0) - critical damping solution:**

$$q_{gf}^{(cd)}(T) = \frac{e^{-\frac{\zeta}{2}T}\left[T\left(-\frac{\zeta}{2}+\frac{1}{\hat{L}}\right)+1\right]}{\hat{r}}, \tag{20}$$

where the characteristic decay time is $\tau_q = \frac{2}{\zeta}$.

**Case 3 ($\Delta$<0) - underdamping solution:**

$$q_{gf}^{(ud)}(T) = \frac{e^{-\frac{\zeta}{2}T}\left[\omega\cos(\omega T)+\left(-\frac{\zeta}{2}+\frac{1}{\hat{L}}\right)\sin(\omega T)\right]}{\omega\hat{r}}, \tag{21}$$

where the characteristic decay time is $\tau_q = \frac{2}{\zeta}$, and the natural angular frequency is $\omega = \frac{\sqrt{|\Delta|}}{2}$.

For criticality, we set $(\hat{r}, \hat{c}, \hat{L}) = (r_c, c_c, L_c)$, so that:

$$\Delta(r_c, c_c, L_c) = 0. \tag{22}$$

Expanding and summarizing Eq. (22), we have:

$$L_c^2 - 2r_c c_c L_c + r_c^2 c_c^2 - 4r_c^2 c_c L_c = 0. \tag{23}$$

Solving Eq. (23) for $L_c$, we get:

$$\begin{cases} L_{c1} = \left(2r_{c1} + 1 - 2\sqrt{r_{c1}^2 + r_{c1}}\right)r_{c1}c_{c1} \\ L_{c2} = \left(2r_{c2} + 1 + 2\sqrt{r_{c2}^2 + r_{c2}}\right)r_{c2}c_{c2} \end{cases}. \tag{24}$$

Therefore, we obtain the resistance-capacitance-inductance phase space, as shown in Fig. 2a. In the phase space are two critical surfaces ($r_{c1}$, $c_{c1}$, $L_{c1}$) and ($r_{c2}$, $c_{c2}$, $L_{c2}$). When $r_{c1}$=$r_{c2}$ and $c_{c1}$=$c_{c2}$, $L_{c1}$<$L_{c2}$. The two critical surfaces separate this phase space into three regions, i.e. two overdamping regions and one underdamping region. In fixing resistance and capacitance and increasing inductance, the state $(\hat{r}, \hat{c}, \hat{L})$ passes through: the first overdamping region (OD1), the first critical damping surface (CD1), lower underdamping region (lower UD), upper underdamping region (upper UD), the second critical damping surface (CD2), and the second overdamping region (OD2).

We select, for instance, six sets of $(\hat{r}, \hat{c}, \hat{L},)$ values to calculate the corresponding Green functions of the charge time series $q_{gf}(T)$ according to Eqs. (19)-(21), as shown on Fig. 2b. The six sets are A (5, 5, 0.1) in the OD1 region, B (5, 5, ~1.1387) on the CD1 surface, C (5, 5, 10) in the lower UD region, D (5, 5, 100) in the upper UD region, E (5, 5, ~548.8613) on the CD2 surface, and F (5, 5, 700) in the OD2 region, respectively. Information concerning the six selected sets, including $\zeta$, $\eta$, $\Delta$, is also listed in Table 2. When $\hat{L} \leq L_{c1}$ and $\Delta \geq 0$, the time series A and B for $q_{gf}(T)$ decay much faster and without oscillations. However, when $\Delta < 0$, C and D for $q_{gf}(T)$ decay while oscillating around zero with a natural frequency, but $q_{gf}(T)$ decays faster for C than for D. Finally, when $\hat{L} \geq L_{c2}$ and $\Delta \geq 0$, $q_{gf}(T)$ for E and F decay much slower with overshooting below zero and rebounding close to zero. It is important that the series $q_{gf}(T)$ behaves quite differently for different damping conditions.

### 3.3 Relationship between stress drops and voltage fluctuations

We assume a given stress history ($\tau$), which is the simulated stress from a spring-block system, as shown in the upper panel of Fig. 3a. If the ratio of stress-induced voltage to stress is $\beta=1$, the time series also represents the stress-induced voltage. Now, we convolve the stress-induced voltage with a Green function $q_{gf}(T)$, as derived in Section 3.2. Dividing this convolution series by the block capacitance ($c$), we obtain a block voltage series ($V_{SB}$), as shown in the lower panel of Fig. 3a. Note that we discuss all results with dimensionless variables. The different series represents the block voltages of sets A to F, whose parameters are listed in Table 2. At the beginning of those series, a short-term transient state exists during the period 0-2000 time unit. This transient state is ignored in the later analysis.

By taking the first difference of the stress history, as shown in the upper panel of Fig. 3b, we get the stress drops ($\Delta\tau$):

$$\Delta\tau_t = \tau_t - \tau_{t-1}, \tag{25}$$

where $t$ is any time point. On the other hand, we also consider relative voltage fluctuations ($V_{fluc}$), as shown in the lower panel of Fig. 3b, defined as:

$$V_{fluc}^{(t)} = \left|\frac{V_{SB}^{(t)} - V_{SB}^{(t-1)}}{V_{SB}^{(t)}}\right| \cdot 100, \tag{26}$$

where $t$ is any time point. In order to compare the stress drop of an event with its corresponding relative voltage fluctuation, we define the maximal value of $V_{fluc}$ associated to an event:

$$V_{fluc}^{max}(i) = max\{V_{fluc}^t, \quad t(i) \leq t < t(i+1)\}, \tag{27}$$

where $t(i)$ is the occurrence time of the $i^{th}$ event. Figure 3c shows the relationship between the stress drop of an event and its corresponding maximal relative voltage fluctuation. Overall, the stress drops and maximal voltage fluctuations follow a linear relationship, especially for moderate-sized events, due to the linear induced-voltage stress (Eq. (9)). However, there are a scattering of voltage fluctuations of small-sized events for sets A-F. It is because when the interevent times of events larger than a specified size are smaller than the characteristic decay times for sets A-F, the induced voltage time series of a small-sized event is interfered with its past event's voltage series. Increasing a specified event size, the interevent times of events larger than this size are greater than the characteristic times, and it is hard for induced voltages to interfere between two successive events, i.e. a clearly linear relation between the moderate-sized events and their voltage fluctuations. When the event sizes are so large that their voltage time series of sets D-F have much longer waves and decay much more slowly, a nonlinear amplification of voltage fluctuations of large-sized events might appear only for sets D-F. It is expected that small and large stress drops generate small and large voltage fluctuations, respectively. In fact, such a simple relationship may not always hold in the interference of induced voltages by past events and in the presence of complex spatio-temporal dynamics of the interacting blocks. In a heterogeneous multi-blocks COS model, it is possible that relatively small ruptures generate locally relatively large voltage fluctuations, especially for the blocks located in the upper UD, CD2, and OD2 regions of the phase space. This phenomenon would allow us to detect foreshock-induced electric signals, while the foreshocks themselves are below the detection threshold of seismic networks.

## 4. Electrokinetic system: multi-blocks problem

### 4.1 Governing equations

It is easy to expand the previous single-block model to a multi-blocks system, as shown in Fig. 1b. All notations remain the same, except that we add a subscript $k$ relative to each block. Furthermore, the polarization direction ($p_d$) of stress-induced voltages should be considered here, so that Eq. (9) becomes:

$$V_{ink}(\tau_k) = p_{dk}\beta_k\tau_k, \tag{28}$$

where $p_{dk} = \begin{cases} 1, & \tau_{k-1} \geq \tau_{k+1} \\ -1, & \tau_{k-1} \leq \tau_{k+1} \end{cases}$, which is roughly consistent with the observation of Freund's experiments that positive holes flow from more stressed areas to less stressed ones. Note that $p_d$ in the $k^{th}$ block is assigned randomly to 1 or -1 when $\tau_{k-1} = \tau_{k+1}$. The Kirchhoff's voltage law in the $k^{th}$ block gives:

$$\begin{cases} V_{in1} - i_{r1}r_1 - \frac{q_1}{c_1} = 0 \\ V_{ink} - i_{rk}r_k - \frac{q_k}{c_k} + \frac{q_{k-1}}{c_{k-1}} = 0, \quad k = 2 \text{ to } N \end{cases}. \tag{29}$$

For the current-charge relation in the $k^{th}$ block capacitor, we have:

$$i_{ck} = \frac{dq_k}{dt}, \quad k = 1 \text{ to } N. \tag{30}$$

We then write the Kirchhoff's law for the current flowing towards the neighboring blocks or ground:

$$\begin{cases} i_{rk} = I_k + i_{ck} + i_{r(k+1)}, & k = 1 \text{ to } N-1 \\ i_{rN} = I_N + i_{cN} \end{cases}. \tag{31}$$

The equality of the voltages of the $k^{th}$ block resistance, stress-induced voltage, and RL components becomes:

$$\begin{cases} I_1 R_1 + \frac{dI_1}{dt} L_1 = V_{in1} - i_{r1} r_1 \\ I_k R_k + \frac{dI_k}{dt} L_k - I_{k-1} R_{k-1} - \frac{dI_{k-1}}{dt} L_{k-1} = V_{ink} - i_{rk} r_k, & k = 2 \text{ to } N \end{cases}. \tag{32}$$

The multi-blocks unknown variable vector is $\mathbf{G}=[q_k, i_{ck}, i_{rk}, I_k]$ for $k=1$ to $N$, with the initial condition $\mathbf{G}(t=0)=\vec{0}$. Finally, the total voltage of the electrokinetic system is defined as:

$$V_{SB} = \frac{1}{N} \sum_{k=1}^{N} \left( R_k I_k + L_k \frac{dI_k}{dt} \right) = \frac{1}{N} \sum_{k=1}^{N} \frac{q_k}{c_k}. \tag{33}$$

The total voltage $V_{SB}$ is analog to the voltage measured in real field experiments, which we propose to be associated with the leakage of currents into the ground. In the COS model, that voltage is proportional to the sum of the grounded currents ($I_k$) through grounded resistors ($R_k$) and grounded inductors ($L_k$), which is also equal to the voltage due to the charges stored in the block capacitors ($c_k$).

Again, we transform Eqs. (28)-(33) with dimensionless variables, as done in Section 3.2, but the dimensionless variables and parameters with subscript $k$. We introduce a dimensionless parameter for the mechanical-electrical coupling: $\widehat{\beta_k}$ for $k=1$ to $N$. The stress-induced voltage Eq. (28) becomes:

$$\widehat{V_{ink}}(\tau_k) = p_{dk} \widehat{\beta_k} \tau_k, \quad k = 1 \text{ to } N. \tag{35}$$

Equations (29)-(33) now become:

$$\begin{cases} \widehat{V_{in}} - \widehat{i_{r1}} \widehat{r_1} - \frac{q_1}{c_1} = 0 \\ \widehat{V_{ink}} - \widehat{i_{rk}} \widehat{r_k} - \frac{q_k}{c_k} + \frac{q_{k-1}}{c_{k-1}} = 0, \quad k = 2 \text{ to } N \end{cases}, \tag{36}$$

$$\widehat{i_{ck}} = \frac{d\widehat{q_k}}{dT}, \quad k = 1 \text{ to } N, \tag{37}$$

$$\begin{cases} \widehat{\iota_{rk}} = \widehat{I_k} + \widehat{\iota_{ck}} + \widehat{\iota_{r(k+1)}}, & k = 1 \text{ to } N-1 \\ \widehat{\iota_{rN}} = \widehat{I_N} + \widehat{\iota_{cN}} \end{cases}, \tag{38}$$

$$\begin{cases} \widehat{I_1} + \frac{d\widehat{I_1}}{dT}\widehat{L_1} = \widehat{V_{in1}} - \widehat{\iota_{r1}}\hat{r}_1 \\ \widehat{I_k} + \frac{d\widehat{I_k}}{dT}\widehat{L_k} - \alpha_{k-1}\widehat{I_{k-1}} - \alpha_{k-1}\frac{d\widehat{I_{k-1}}}{dT}\widehat{L_{k-1}} = \widehat{V_{ink}} - \widehat{\iota_{rk}}\hat{r}_k, & k = 2 \text{ to } N \end{cases}, \tag{39}$$

$$\widehat{V_{SB}} = \frac{1}{N}\sum_{k=1}^{N}\left(\widehat{I_k} + \widehat{L_k}\frac{d\widehat{I_k}}{dT}\right) = \frac{1}{N}\sum_{k=1}^{N}\frac{\widehat{q_k}}{\widehat{c_k}}. \tag{40}$$

In Eq. (39) $\alpha$ is the ratio of adjacent grounded resistances ($\alpha_k=R_k/R_{k+1}$), which we set to unity, meaning that the grounded resistance is spatially homogeneous. For the sake of simplicity, we set the same parameters $[\widehat{\beta_k},\hat{r}_k,\hat{c}_k,\widehat{L_k}]$ for all blocks, i.e. $[\hat{\beta},\hat{r},\hat{c},\hat{L}]$, so that all blocks belong to the same region of the phase space analyzed in Section 3.2. We also set $\hat{\beta} = 1$, as it will act as a simple scaling factor for all voltages, and we are left to study the effects of varying $\hat{r}$, $\hat{c}$, and $\hat{L}$ on the simulated voltage $\widehat{V_{SB}}$.

## 4.2 Results of the numerical analysis

In the COS model, the multi-blocks problem is far more complicated than the single block one due to interactions between elements. Hence, we solve all the differential equations of the mechanical and electrokinetic systems numerically (using a 4$^{th}$ order Runge-Kutta method). We first implement the spring-block system in order to generate the stress ($\tau_k$) acting on each block and use Eq. (35) to generate the stress-induced voltage ($\widehat{V_{ink}}$). We simulate the fully coupled COS model using the values of electrokinetic parameters listed in Table 2. In these simulations, the number of blocks is $N=128$, and the stiffness ratio is $s=30$, as shown in Table 1.

We observe how the simulated voltage behaves at different damping conditions, as shown in Fig. 4. Figure 4a shows the simulated slips ($D_{SB}$), simulated voltage ($\widehat{V_{SB}}$), and relative voltage fluctuations ($V_{fluc}$) defined as in Eq. (26). Repeating the same analysis of correlations between slips and relative voltage fluctuations of Section 3.3 on the multi-blocks case, we get the empirical relationship between the slip during an event and its corresponding maximal $V_{fluc}$ value (defined as Eq. (27)), as shown in Fig. 4b. The scatterplot of Fig. 4b is separated into four quadrants (Q1 to Q4), whose boundaries are naturally identified by the gaps observed in the distributions of slip and voltage fluctuations ($D_{SB}^{thr} = 1$ sets the vertical threshold separation, $V_{fluc}^{thr} = 100$ sets the horizontal one). For relatively small slips, we consider the ratios $\mathcal{Q}_1 = Q1/(Q1 + Q2)$ and $\mathcal{Q}_2 = Q2/(Q1 + Q2)$, respectively. Quadrant Q1 features events with small slips but large voltage fluctuations, while Q2 features events with small slips and small voltage fluctuations. We observe that $\mathcal{Q}_2 > \mathcal{Q}_1$, i.e. most of small events generate small voltage variations. For relatively large slips, we compute $\mathcal{Q}_3 = Q3/(Q3 + Q4)$ as well as $\mathcal{Q}_4 = Q4/(Q3 + Q4)$. It is $\mathcal{Q}_4 > \mathcal{Q}_3$ that intriguingly most of large events cannot generate large voltage variations. This suggests an explanation for why it is difficult to find out large voltage fluctuations during a strong earthquake in nature.

The ratios are listed in Table 2. In Fig. 4c, one can observe that the ratios defined above for cases A to C are different than for cases D to F. This suggests that there exists a transition of slip-voltage relationships between the lower and upper UD regions as defined in the phase space on Fig. 2a. This predicts a possible variability of slip-induced voltage statistics depending on local constitutive parameters. This variability may explain in turn why large earthquake slips are not systematically followed by large electric signals, as the crust is not in an electrokinetic damping state favorable to such dynamics. On the other hand, our results clearly suggest that precursory electromagnetic signals may be observed before large events if: (i) there are slip foreshocks, i.e. small earthquakes that would be too small to be detected seismically; (ii) the local electrokinetic damping conditions allow them to leave a measurable electromagnetic fingerprint. The COS model thus offers a nice opportunity to test for this feature, provided the spring-block model is modified in order to allow for such small precursory slips (recall that standard Burridge-Knopoff models contain no or very rare foreshocks (Pepke & Carlson, 1994)).

## 5. Discussion

Field observations of electromagnetic signals suggest the existence of propagating unipolar pulses prior to earthquakes (Nenovski, 2016; Scoville et al., 2015; Tsai et al., 2006). The proposed COS model also generates unipolar voltage changes due to local stress drops (especially for E and F in Fig. 2), which could be analogous to the real observations. Furthermore, the small-scale ruptures before a large event could generate unipolar signals with different properties, depending on the underground electrokinetic parameters, their amplitudes and shapes being controlled by the underground resistance, capacitance, and inductance. Besides, the background values of the spring-block voltage $\widehat{V_{SB}}$ is not zero (see Figs. 3a and 4a), suggesting that the measurement of mean values of natural occurring geoelectric fields in a certain period might be used to infer to the stress level of the region. Hence, it would be possible to use the geoelectric field to invert for the stress level.

Relationships between the geoelectric field skewness and kurtosis, on the one hand, and earthquakes, on the other hand, have been recently reported (Chen et al., 2017; Chen & Chen, 2016; see our other article presenting earthquake forecasting based on geoelectric data in this special volume), suggesting that the statistical distribution of amplitudes of geoelectric signals is modified during the preparation stage of earthquakes. Figure 5 shows the time series of the event slips, as well as the skewness and kurtosis of the $\widehat{V_{SB}}$ series for the multi-blocks COS model described above. Using a moving window technique, we calculate the skewness and kurtosis within a window length of 131 time units, which is the median of inter-event times of events with $D_{SB}>1$. It seems that slips, and even micro-slips, perturb electric signals, as both skewness and kurtosis time series display quite ample fluctuations. The proposed COS model thus also provides an explanation for similar transients observed in real systems. We leave for another study the detailed analysis of these electric signals as predictors of the sliding events in the COS model.

Eftaxias et al. (2003) have performed power spectrum analyses of electromagnetic signals before, during, and after a large earthquake, suggesting an increase of the low-frequency energy content, as well as a power-law-shaped spectrum prior to large events. If the underlying mechanism of the electromagnetic signals obeys critical dynamics, then its spectrum is expected to behave as $S(f) = \frac{a}{f^b}$, that is $log(S) = log(a) - b \cdot log(f)$. The exponent $b$ of the power-law spectrum in field observations is observed to become closer to 2 during the pre-seismic critical stage, a value separating regimes of anti-persistent and persistent behavior of the electromagnetic time series. From the COS model's view point, the slopes of the power-law fit to the power spectra depend on the damping conditions of the underground electrokinetic parameters, as shown in Fig. 6. To construct Fig. 6, we consider a full voltage time series excluding its transient initial state, and divide it into 305 non-overlapping segments, each of duration of 131 time units as above. We estimate the power spectrum and its exponent $b$ for each time window by fitting $log(S) = log(a) - b \cdot log(f)$ with a least square method. Analyzing statistically the 305 power spectra and $b$ values, we get the average of the power spectra (Fig. 6a) as well as the statistics of their slopes (Fig. 6b) for sets A to F. It seems that, once again, there is a transition between the lower and upper UD regions of the phase space (see also the last column of Table 2), which is similar to the result of Eftaxias et al. (2003).

A possible interpretation is that the pre-seismic critical transition in Eftaxias et al. (2003)'s study might be caused by the changes of the underground electrokinetic parameters during an earthquake preparation process.

Moreover, Potirakis et al. (2013) emphasized that the pre-seismic electromagnetic emissions are due to the progressive fracturing of the heterogeneous system that surrounds the main fault. However, in our study, even a homogeneous system can also produce anomalous and complex voltages depending on the state of the electrokinetic parameters. This suggests that heterogeneity of a system is not necessary to produce the complicated fracture-induced electromagnetic emissions prior to large earthquakes. The stress changes and interferences of induced voltage series of two successive events indeed appear to be the key to produce geoelectric variations.

However, future studies might also focus on stress-induced damage and fluid flow, which would make underground mechano-electrokinetic parameters dependent on space and time as well.

# Acknowledgements

HJC and CCC are supported by Grant No. MOTC-CWB-107-E-01 from Taiwan Central Weather Bureau and by Grant No. MOST-106-2116-M-008-002 from Taiwan Ministry of Science and Technology. The authors would like to thank the anonymous reviewers and the editor for their valuable comments and suggestions.

# References

Abaimov, S. G., Turcotte, D. L., Shcherbakov, R., & Rundle, J. B. (2007). Recurrence and interoccurrence behavior of self-organized complex phenomena. *Nonlinear Processes in Geophysics*, *14*(4), 455–464.

Andersen, J. V., Sornette, D., & Leung, K. (1997). Tricritical Behavior in Rupture Induced by Disorder. *Physical Review Letters*, *78*(11), 2140–2143. https://doi.org/10.1103/PhysRevLett.78.2140

Bleier, T., Dunson, C., Maniscalco, M., Bryant, N., Bambery, R., & Freund, F. (2009). Investigation of ULF magnetic pulsations, air conductivity changes, and infra red signatures associated with the 30 October Alum Rock M5.4 earthquake. *Natural Hazards and Earth System Science*, *9*(2), 585–603. https://doi.org/10.5194/nhess-9-585-2009

Brown, S. R., Scholz, C. H., & Rundle, J. B. (1991). A simplified spring-block model of earthquakes. *Geophysical Research Letters*, *18*(2), 215–218. https://doi.org/10.1029/91GL00210

Burridge, R., & Knopoff, L. (1967). Model and theoretical seismicity. *Bulletin of the Seismological Society of America*, *57*(3), 341–371.

Cao, T., & Aki, K. (1984). Seismicity simulation with a mass-spring model and a displacement hardening-softening friction law. *Pure and Applied Geophysics*, *122*(1), 10–24. https://doi.org/10.1007/BF00879646

Carlson, J. M. (1991). Two-dimensional model of a fault. *Physical Review A*, *44*(10), 6226–6232. https://doi.org/10.1103/PhysRevA.44.6226

Carlson, J. M., & Langer, J. S. (1989a). Mechanical model of an earthquake fault. *Physical Review A*, *40*(11), 6470–6484. https://doi.org/10.1103/PhysRevA.40.6470

Carlson, J. M., & Langer, J. S. (1989b). Properties of earthquakes generated by fault dynamics. *Physical Review Letters*, *62*(22), 2632–2635. https://doi.org/10.1103/PhysRevLett.62.2632

Carlson, J. M., Langer, J. S., Shaw, B. E., & Tang, C. (1991). Intrinsic properties of a Burridge-Knopoff model of an earthquake fault. *Physical Review A*, *44*(2), 884–897. https://doi.org/10.1103/PhysRevA.44.884

Carlson, J. M., Langer, J. S., & Shaw, B. E. (1994). Dynamics of earthquake faults. *Reviews of Modern Physics*, *66*(2), 657–670. https://doi.org/10.1103/RevModPhys.66.657

Cartwright, J. H. E., Hernández-García, E., & Piro, O. (1997). Burridge-Knopoff models as elastic excitable media. *Physical Review Letters*, *79*(3), 527–530. https://doi.org/10.1103/PhysRevLett.79.527

Chen, C.-C., & Wang, J.-H. (2010). One-dimensional dynamical modeling of slip pulses. *Tectonophysics*, *487*(1–4), 100–104. https://doi.org/10.1016/j.tecto.2010.03.010

Chen, C.-C., Wang, J.-H., & Huang, W.-J. (2012). Material decoupling as a mechanism of aftershock generation. *Tectonophysics*, *546–547*, 56–59. https://doi.org/10.1016/j.tecto.2012.04.016

Chen, H.-J., & Chen, C.-C. (2016). Testing the correlations between anomalies of statistical indexes of the geoelectric system and earthquakes. *Natural Hazards*, *84*(2), 877–895. https://doi.org/10.1007/s11069-016-2460-4

Chen, H.-J., Chen, C.-C., Ouillon, G., & Sornette, D. (2017). Using skewness and kurtosis of geoelectric fields to forecast the 2016/2/6, ML6.6 Meinong, Taiwan Earthquake. *Terrestrial, Atmospheric and Oceanic Sciences*, *28*(5), 745–761. https://doi.org/10.3319/TAO.2016.11.01.01

Christensen, K., & Olami, Z. (1992a). Scaling, phase transitions, and nonuniversality in a self-organized critical cellular-automaton model. *Physical Review A*, *46*(4), 1829–1838. https://doi.org/10.1103/PhysRevA.46.1829

Christensen, K., & Olami, Z. (1992b). Variation of the Gutenberg-Richter b values and nontrivial temporal correlations in a spring-block model for earthquakes. *Journal of Geophysical Research*, *97*(B6), 8729–8735. https://doi.org/10.1029/92JB00427

Eftaxias, K., Kapiris, P., Polygiannakis, J., Bogris, N., Kopanas, J., Antonopoulos, G., et al. (2001). Signature of pending earthquake from electromagnetic anomalies. *Geophysical Research Letters*, *28*(17), 3321–3324. https://doi.org/10.1029/2001GL013124

Eftaxias, K., Kapiris, P., Polygiannakis, J., Peratzakis, A., Kopanas, J., Antonopoulos, G., & Rigas, D. (2003). Experience of short term earthquake precursors with VLF–VHF electromagnetic emissions. *Natural Hazards and Earth System Sciences*, *3*(3/4), 217–228. https://doi.org/10.5194/nhess-3-217-2003

Erickson, B., Birnir, B., & Lavallée, D. (2008). A model for aperiodicity in earthquakes. *Nonlinear Processes in Geophysics*, *15*(1), 1–12. https://doi.org/10.5194/npg-15-1-2008

Erickson, B., Birnir, B., & Lavallée, D. (2011). Periodicity, chaos and localization in a Burridge–Knopoff model of an earthquake with rate-and-state friction. *Geophysical Journal International*, *187*(1), 178–198. https://doi.org/10.1111/j.1365-246X.2011.05123.x

Freund, F. (2003). Rocks that crackle and sparkle and glow : Strange pre-earthquake phenomena. *Journal of Scientific Exploration*, *17*(1), 37–71.

Freund, F. (2007). Pre-earthquake signals – Part I: Deviatoric stresses turn rocks into a source of electric currents. *Natural Hazards and Earth System Sciences*, *7*(5), 535–541. https://doi.org/10.5194/nhess-7-535-2007

Freund, F., & Pilorz, S. (2012). Electric currents in the Earth crust and the generation of pre-earthquake ULF signals. *Frontier of Earthquake Prediction Studies*, 464–508.

Freund, F., & Sornette, D. (2007). Electro-magnetic earthquake bursts and critical rupture of peroxy bond networks in rocks. *Tectonophysics*, *431*(1–4), 33–47. https://doi.org/10.1016/j.tecto.2006.05.032

Freund, F., Kulahci, I. G., Cyr, G., Ling, J., Winnick, M., Tregloan-Reed, J., & Freund, M. M. (2009). Air ionization at rock surfaces and pre-earthquake signals. *Journal of Atmospheric and Solar-Terrestrial Physics*, *71*(17–18), 1824–1834. https://doi.org/10.1016/j.jastp.2009.07.013

Hasumi, T., Chen, C., Akimoto, T., & Aizawa, Y. (2010). The Weibull–log Weibull transition of interoccurrence time for synthetic and natural earthquakes. *Tectonophysics*, *485*(1–4), 9–16. https://doi.org/10.1016/j.tecto.2009.11.012

Ikeya, M., Yamanaka, C., Mattsuda, T., Sasaoka, H., Ochiai, H., Huang, Q., et al. (2000). Electromagnetic pulses generated by compression of granitic rocks and animal behavior. *Episodes*, *23*(4), 262–265.

Ishido, T., & Mizutani, H. (1981). Experimental and theoretical basis of electrokinetic phenomena in rock-water systems and its applications to geophysics. *Journal*

*of Geophysical Research: Solid Earth*, *86*(B3), 1763–1775. https://doi.org/10.1029/JB086iB03p01763

Mitsui, Y., & Cocco, M. (2010). The role of porosity evolution and fluid flow in frictional instabilities: A parametric study using a spring-slider dynamic system. *Geophysical Research Letters*, *37*(23), L23305. https://doi.org/10.1029/2010GL045672

Mizutani, H., Ishido, T., Yokokura, T., & Ohnishi, S. (1976). Electrokinetic phenomena associated with earthquakes. *Geophysical Research Letters*, *3*(7), 365–368. https://doi.org/10.1029/GL003i007p00365

Nabawy, B. S. (2015). Impacts of the pore- and petro-fabrics on porosity exponent and lithology factor of Archie's equation for carbonate rocks. *Journal of African Earth Sciences*, *108*, 101–114. https://doi.org/10.1016/j.jafrearsci.2015.04.014

Nenovski, P. (2016). Unipolar magnetic field pulses as transient signals prior to the 2009 Aquila earthquake shock. *ArXiv:1602.02985 [Physics]*. Retrieved from http://arxiv.org/abs/1602.02985

Nitsan, U. (1977). Electromagnetic emission accompanying fracture of quartz-bearing rocks. *Geophysical Research Letters*, *4*(8), 333–336. https://doi.org/10.1029/GL004i008p00333

Nussbaum, J., & Ruina, A. (1987). A two degree-of-freedom earthquake model with static/dynamic friction. *Pure and Applied Geophysics*, *125*(4), 629–656. https://doi.org/10.1007/BF00879576

Pepke, S. L., & Carlson, J. M. (1994). Predictability of self-organizing systems. *Physical Review E*, *50*(1), 236–242. https://doi.org/10.1103/PhysRevE.50.236

Potirakis, S. M., Karadimitrakis, A., & Eftaxias, K. (2013). Natural time analysis of critical phenomena: The case of pre-fracture electromagnetic emissions. *Chaos: An Interdisciplinary Journal of Nonlinear Science*, *23*(2), 023117. https://doi.org/10.1063/1.4807908

Sasaoka, H., Yamanaka, C., & Ikeya, M. (1998). Measurements of electric potential variation by piezoelectricity of granite. *Geophysical Research Letters*, *25*(12), 2225–2228. https://doi.org/10.1029/98GL51179

Scoville, J., Heraud, J., & Freund, F. (2015). Pre-earthquake magnetic pulses. *Nat.*

*Hazards Earth Syst. Sci.*, *15*(8), 1873–1880. https://doi.org/10.5194/nhess-15-1873-2015

Sornette, A., & Sornette, D. (1990). Earthquake rupture as a critical point: consequences for telluric precursors. *Tectonophysics*, *179*(3–4), 327–334. https://doi.org/10.1016/0040-1951(90)90298-M

Stavrakas, I., Triantis, D., Agioutantis, Z., Maurigiannakis, S., Saltas, V., Vallianatos, F., & Clarke, M. (2004). Pressure stimulated currents in rocks and their correlation with mechanical properties. *Natural Hazards and Earth System Sciences*, *4*(4), 563–567. https://doi.org/10.5194/nhess-4-563-2004

Takeuchi, A., Lau, B. W. S., & Freund, F. T. (2006). Current and surface potential induced by stress-activated positive holes in igneous rocks. *Physics and Chemistry of the Earth, Parts A/B/C*, *31*(4–9), 240–247. https://doi.org/10.1016/j.pce.2006.02.022

Telesca, L., Colangelo, G., Lapenna, V., & Macchiato, M. (2004). Fluctuation dynamics in geoelectrical data: an investigation by using multifractal detrended fluctuation analysis. *Physics Letters A*, *332*(5–6), 398–404. https://doi.org/10.1016/j.physleta.2004.10.011

Telesca, L., Lovallo, M., Romano, G., Konstantinou, K. I., Hsu, H.-L., & Chen, C. (2014). Using the informational Fisher–Shannon method to investigate the influence of long-term deformation processes on geoelectrical signals: An example from the Taiwan orogeny. *Physica A: Statistical Mechanics and Its Applications*, *414*, 340–351. https://doi.org/10.1016/j.physa.2014.07.060

Tsai, Y.-B., Liu, J.-Y., Ma, K.-F., Yen, H.-Y., Chen, K.-S., Chen, Y.-I., & Lee, C.-P. (2006). Precursory phenomena associated with the 1999 Chi-Chi earthquake in Taiwan as identified under the iSTEP program. *Physics and Chemistry of the Earth, Parts A/B/C*, *31*(4–9), 365–377. https://doi.org/10.1016/j.pce.2006.02.035

Vallianatos, F., & Triantis, D. (2008). Scaling in Pressure Stimulated Currents related with rock fracture. *Physica A: Statistical Mechanics and Its Applications*, *387*(19–20), 4940–4946. https://doi.org/10.1016/j.physa.2008.03.028

Varotsos, P., & Alexopoulos, K. (1984). Physical properties of the variations of the electric field of the earth preceding earthquakes. II. determination of epicenter and magnitude. *Tectonophysics*, *110*(1–2), 99–125.

https://doi.org/10.1016/0040-1951(84)90060-X

Varotsos, Panayiotis. (2005). *The Physics of Seismic Electric Signals*. Tokyo: TERRAPUB, Terra Scientific Publishing.

Wang, J. (2012). Some intrinsic properties of the two-dimensional dynamical spring-slider model of earthquake faults. *Bulletin of the Seismological Society of America*, *102*(2), 822–835. https://doi.org/10.1785/0120110172

Wang, J. H. (2008). One-dimensional dynamical modeling of earthquakes: A review. *Terrestrial, Atmospheric and Oceanic Sciences*, *19*, 183–203. https://doi.org/10.3319/TAO.2008.19.3.183(T)

Wang, Jeen-Hwa. (2009). A numerical study of comparison of two one-state-variable, rate- and state-dependent friction evolution laws. *Earthquake Science*, *22*(2), 197–204. https://doi.org/10.1007/s11589-009-0197-9

Wang, J.-H., & Hwang, R.-D. (2001). One-dimensional dynamic simulations of slip complexity of earthquake faults. *Earth, Planets and Space*, *53*(2), 91–100.

Yamada, I., Masuda, K., & Mizutani, H. (1989). Electromagnetic and acoustic emission associated with rock fracture. *Physics of the Earth and Planetary Interiors*, *57*(1–2), 157–168. https://doi.org/10.1016/0031-9201(89)90225-2

Yoshida, S., & Kato, N. (2003). Episodic aseismic slip in a two-degree-of-freedom block-spring model. *Geophysical Research Letters*, *30*(13), 1681. https://doi.org/10.1029/2003GL017439

# Figures and Figure Captions

(a)

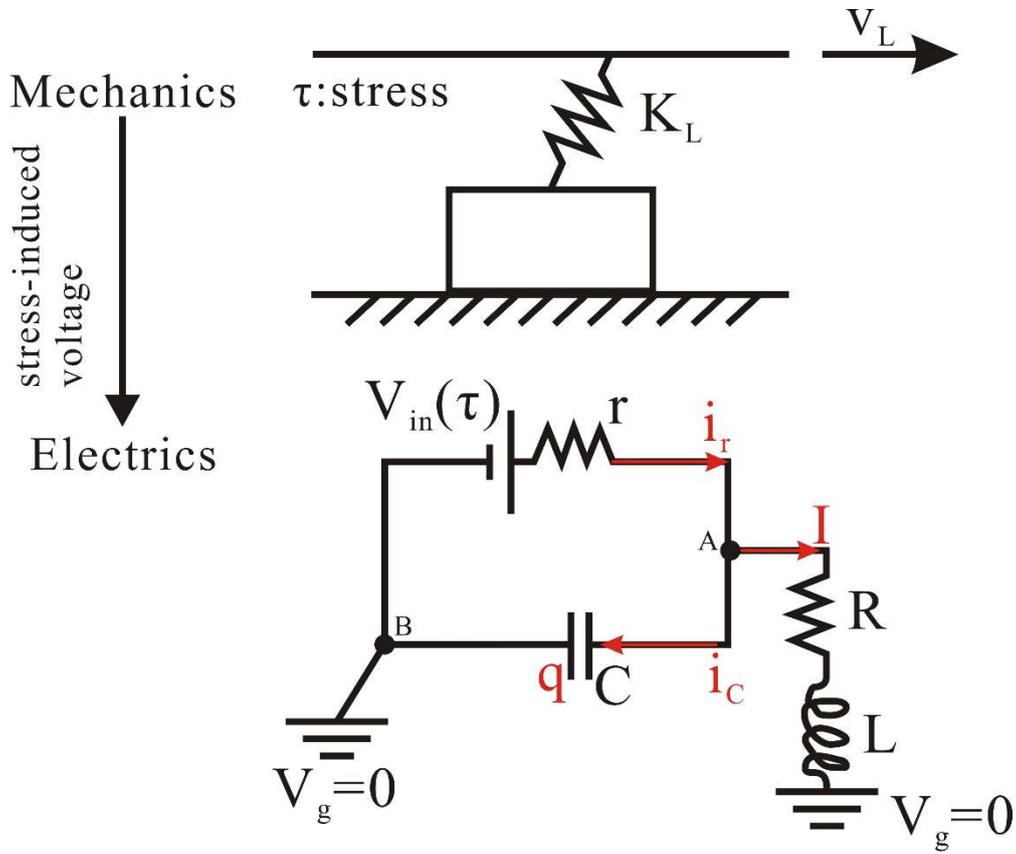

(b)

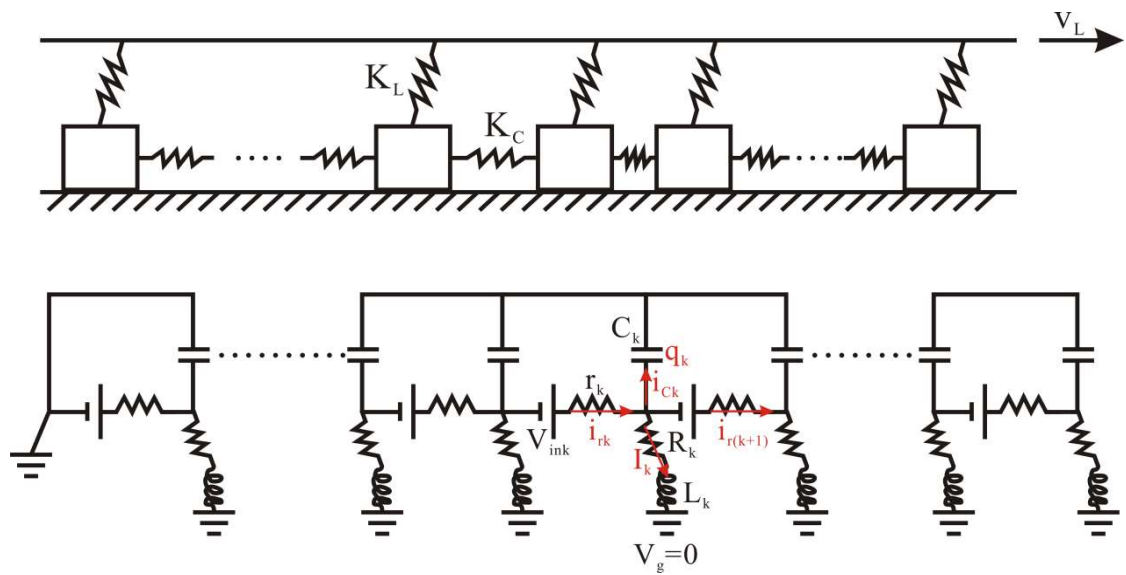

Figure 1. Schematic diagram of the Chen-Ouillon-Sornette model, combining a spring-block system and an electrokinetic system (a) for a single-block case, and (b) for a multi-blocks case.

(a)

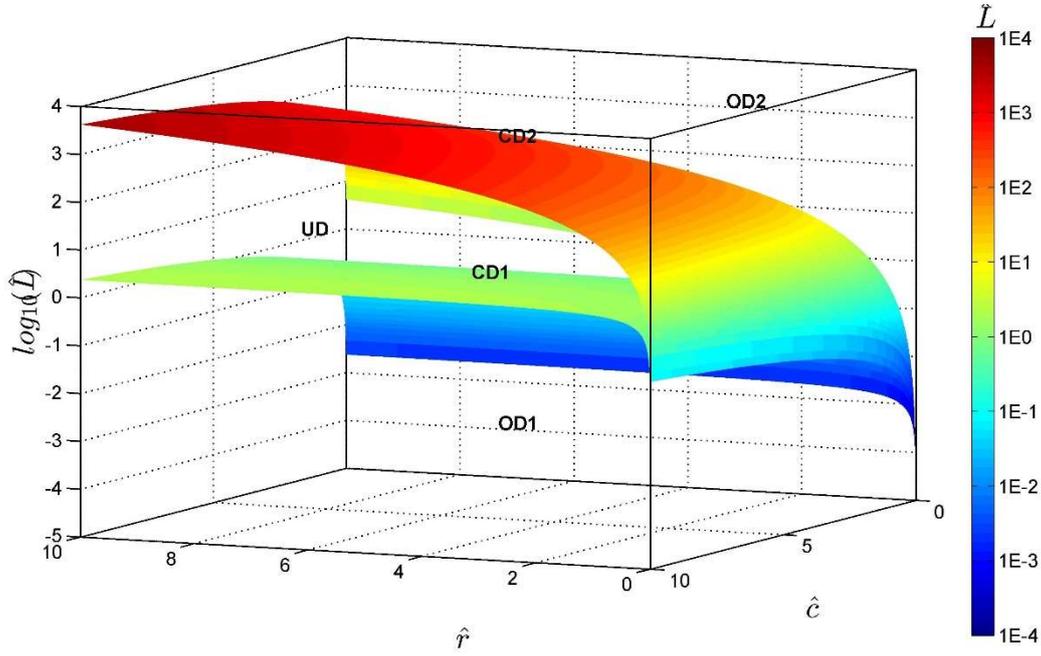

(b)

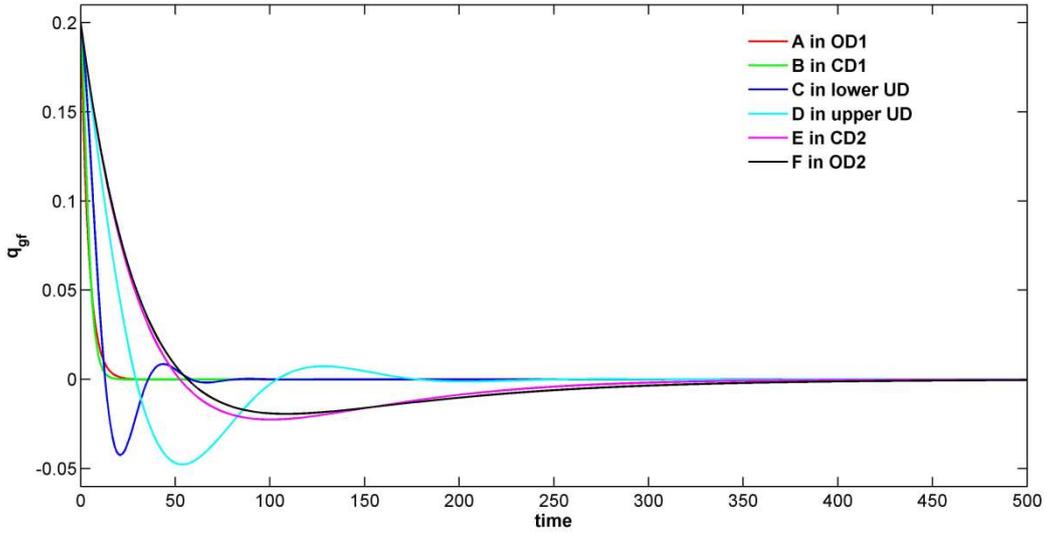

Figure 2. (a) Resistance-capacitance-inductance phase space, and (b) Green functions of charge time series $q_{gf}(T)$ corresponding to different damping regions of the phase space. Information concerning sets A to F is listed in Table 2. The names of each regions and separating surfaces are defined in the main text: OD1 (first over damping region); CD1 (first critical damping surface); UD (underdamping region); CD2 (second critical damping surface); OD2 (second overdamping region).

(a)

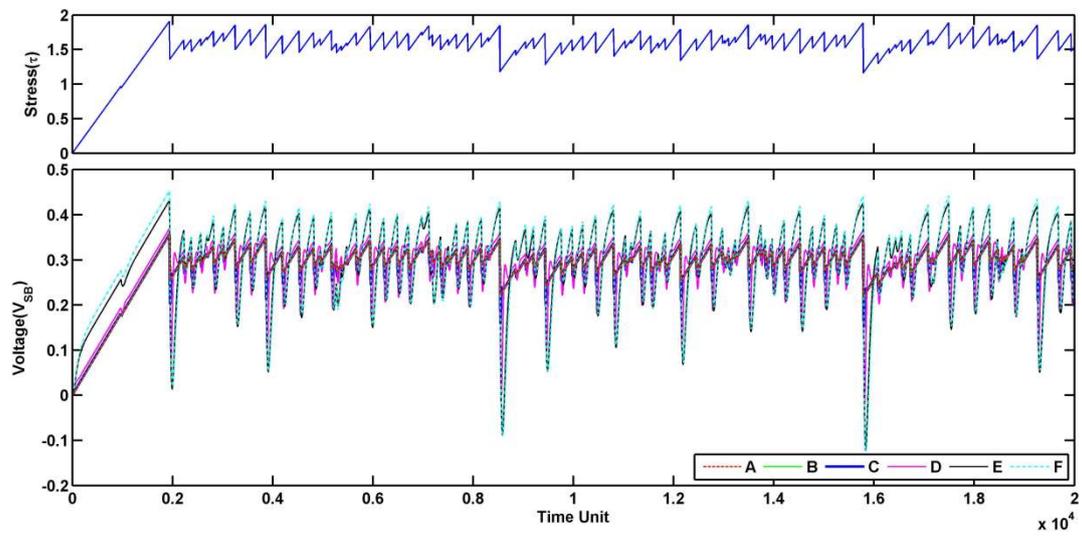

(b)

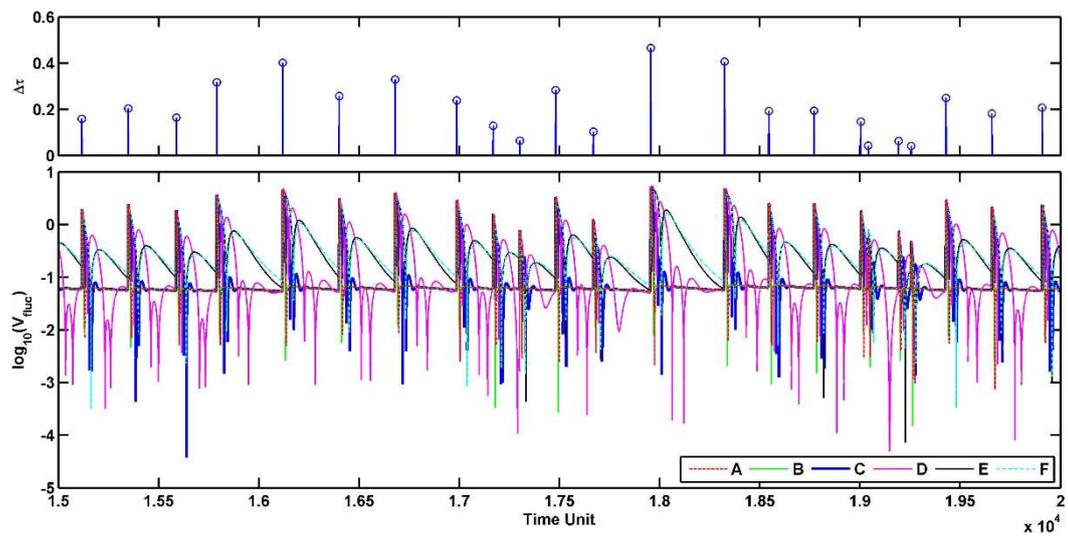

(c)

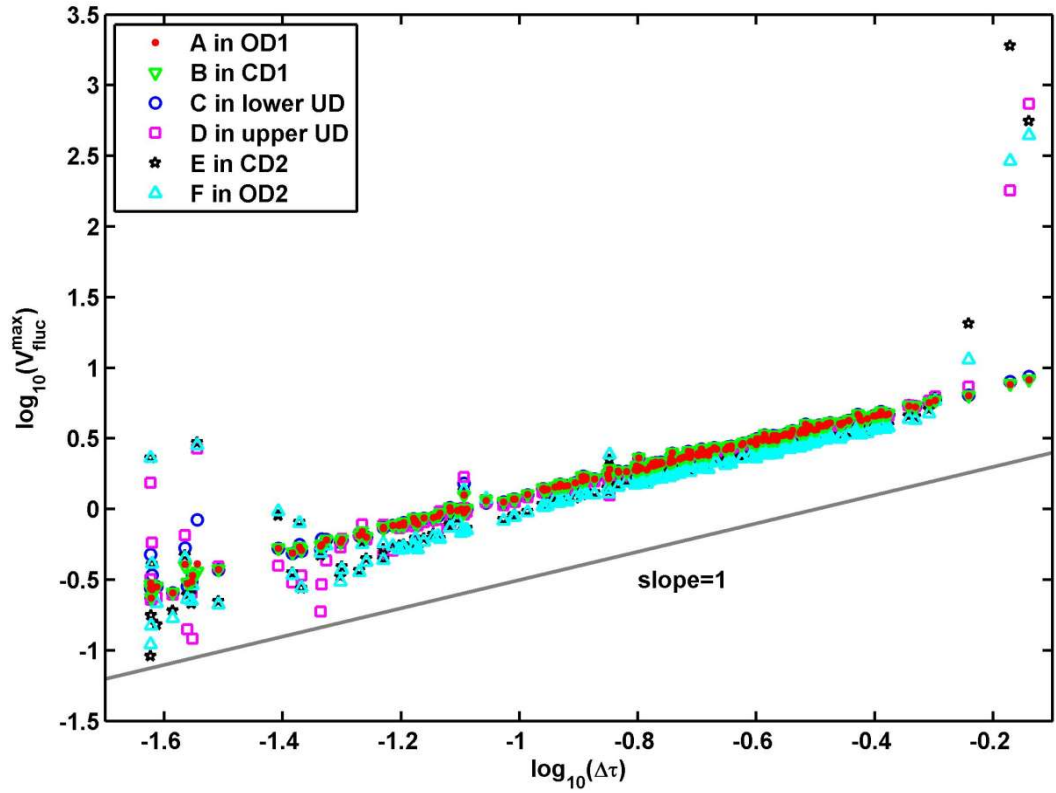

Figure 3. In the single-block model, (a) time series of stress ($\tau$) and its block voltage ($V_{SB}$) for sets A to F. (b) Time series of stress drops ($\Delta\tau$) of events and relative voltage fluctuation ($V_{fluc}$) for sets A to F. (c) Scatter plot of the stress drop of an event and its corresponding maximal voltage fluctuation in a log-log scale. The gray line with slope=1 is plotted as reference.

(a)

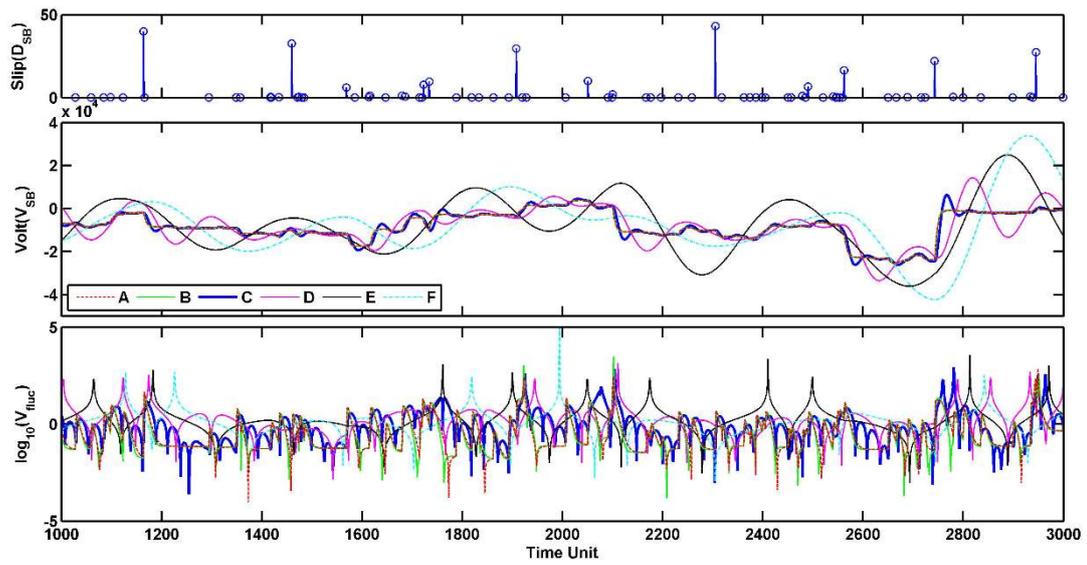

(b)

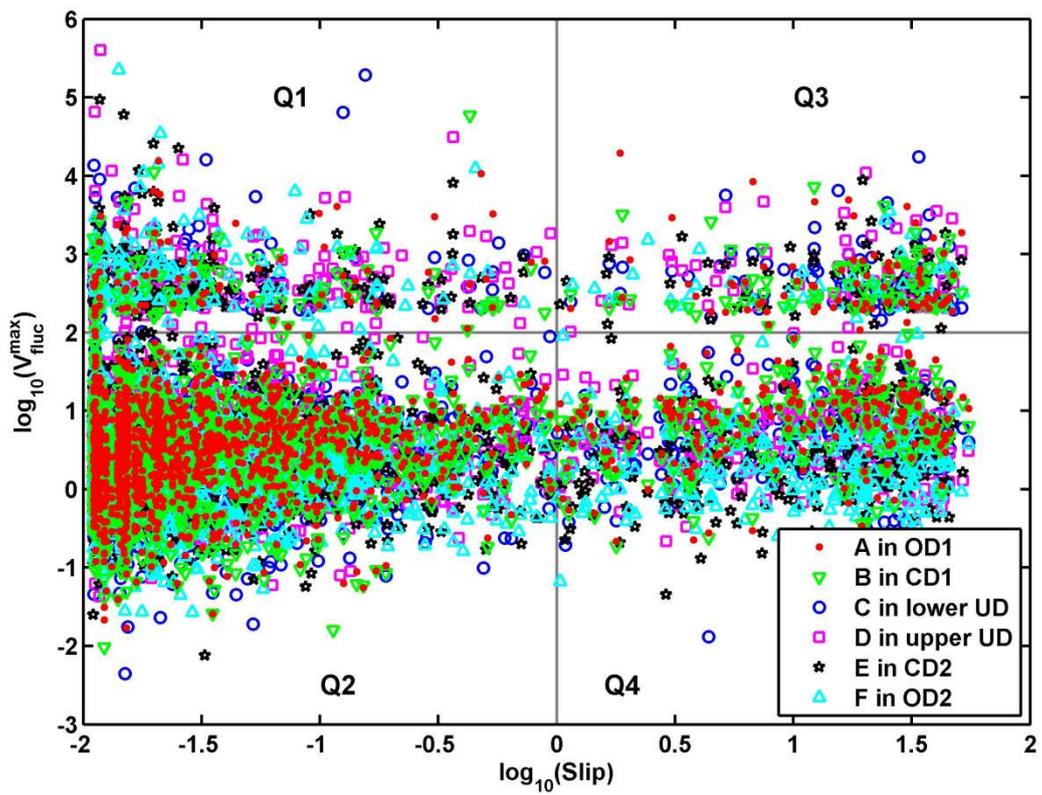

(c)

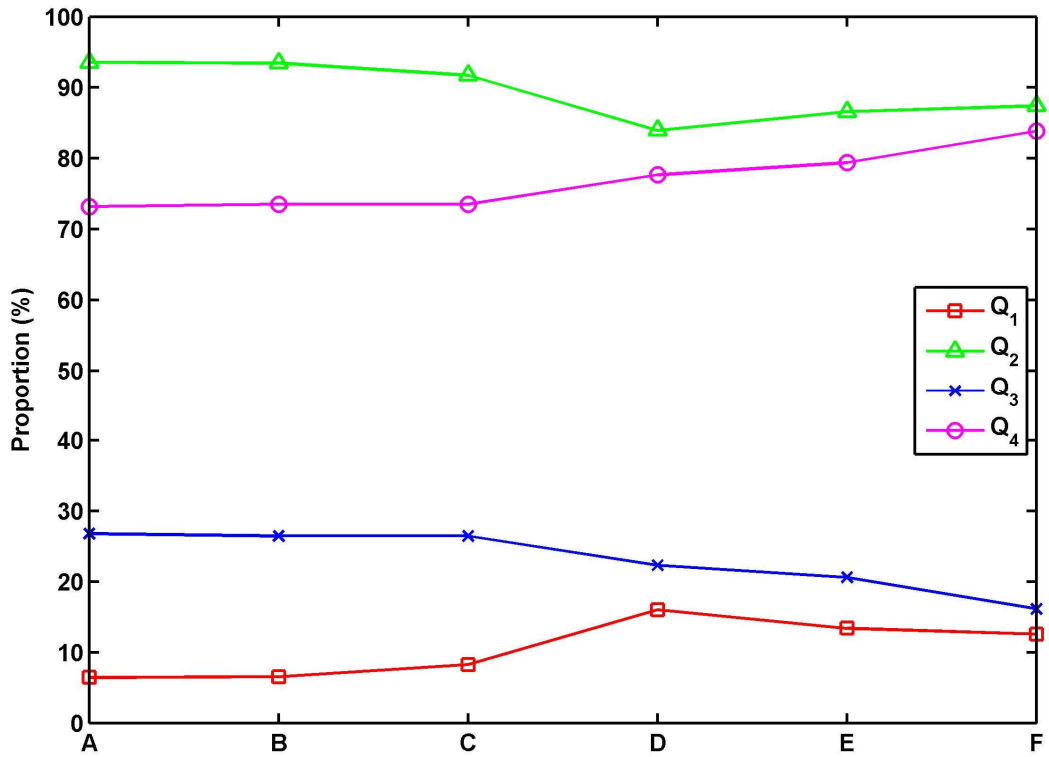

Figure 4. In the multi-block model, (a) time series of event slips ($D_{SB}$) and block voltage ($V_{SB}$) and relative voltage fluctuations ($V_{fluc}$) for sets A to F. (b) Scatter plot of the slip amplitude of an event and its corresponding maximal voltage fluctuation. (c) Proportions of small or large voltage fluctuations corresponding to small or large slips for sets A to F. Note that $Q_1+Q_2$=100% and $Q_3+Q_4$=100%, respectively.

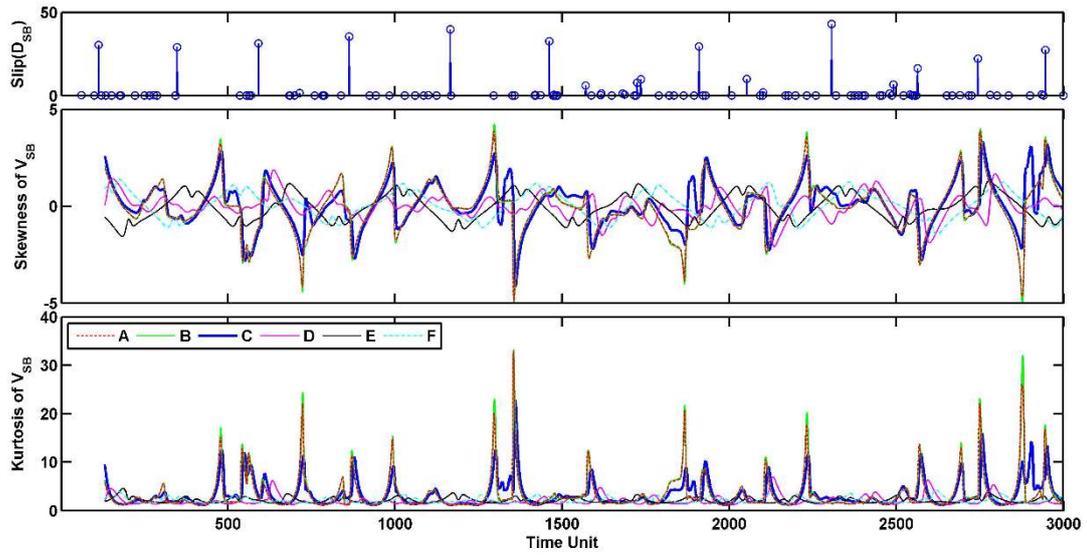

Figure 5. Time series of slips, skewness, and kurtosis for sets A to F. Blue circles give the amount of slip of one event. Skewness and kurtosis are calculated in a moving window length of 131 time units, which is the median of inter-event times of events with $D_{SB}$>1.

(a)

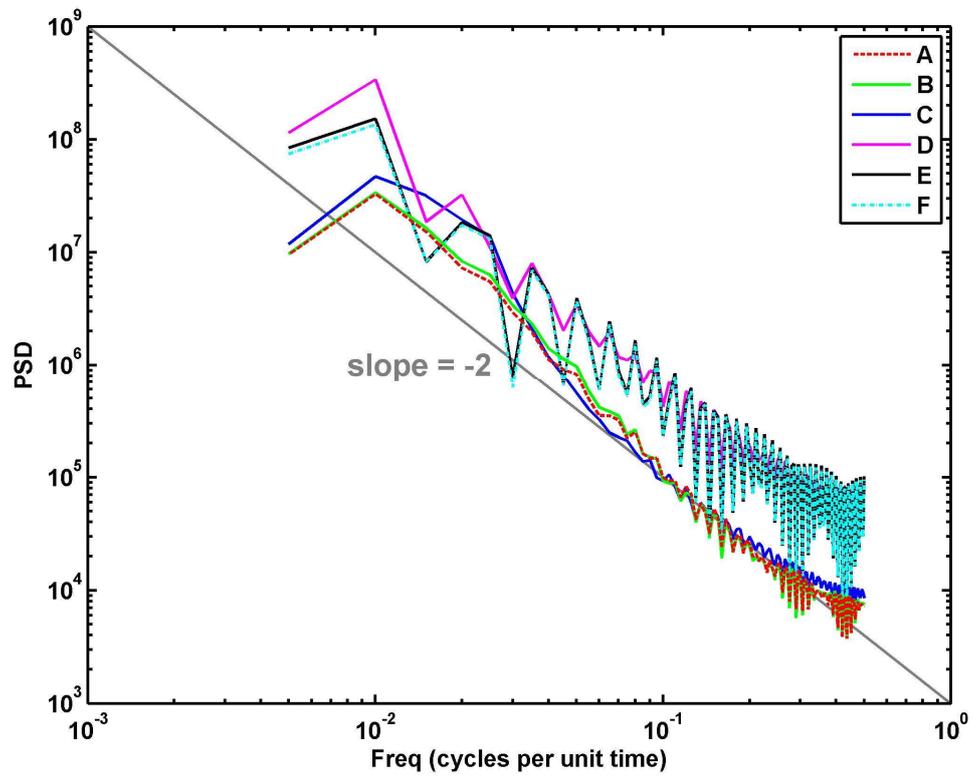

(b)

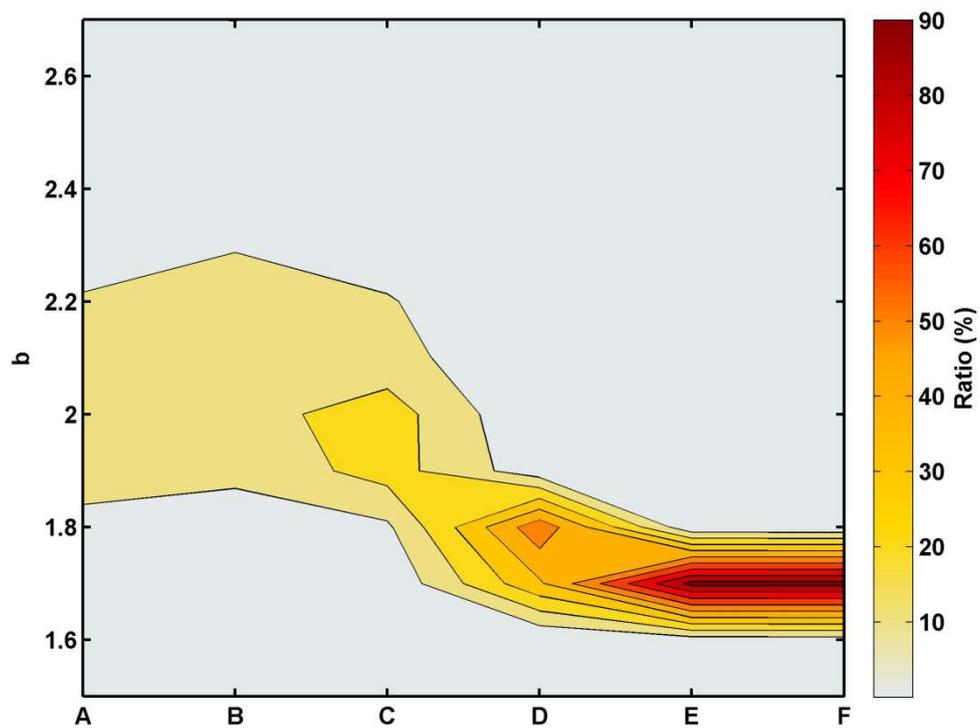

Figure 6. (a) Average of power spectrum densities (PSDs) of the 305 non-overlapping segments, each segment of length equal to 131 time units. The gray line with slope=-2 is plotted as reference. (b) Statistical distributions of power-law exponents ($b$ value) of obtained by calibration of the power law in each of the 305 PSDs.

# Tables and Table Captions

Table 1. Descriptions and values of the Chen-Ouillon-Sornette model's parameters.

| Symbol | Description | Value |
|---:|---|---|
| $N$ | Number of the blocks | 128 |
| $K_C$ | Spring constant between adjacent blocks | |
| $K_L$ | Spring constant between one block to the loading plate | |
| $s$ | Stiffness ratio of $K_C$ to $K_L$ | 30 |
| $\phi$ | Ratio of static to dynamic friction | 1.5 |
| $\mu$ | Ratio of limiting static to reference static friction | (1, 3.5) |
| $v_L$ | Velocity of the loading plate | |
| $\tau$ | Dimensionless stress of one block | |
| $D_{SB}$ | Dimensionless slip of the whole spring-block system | |
| $V_{in}$ ($\widehat{V_{in}}$) | (Dimensionless) Stress-induced voltage | $V_{in} = \beta \cdot \tau$ |
| $p_d$ | Polarization direction of stress-induced voltage | {-1,1} |
| $\beta$ ($\hat{\beta}$) | (Dimensionless) Ratio of stress-induced voltage to stress | $\hat{\beta} = 1$ |
| $r$ ($\hat{r}$) | (Dimensionless) Resistance of a block | Listed in Table 2 |
| $R$ | Resistance for charge propagation to ground | |
| $c$ ($\hat{c}$) | (Dimensionless) Capacitance of a block | Listed in Table 2 |
| $L$ ($\hat{L}$) | (Dimensionless) Inductance for charge propagation to ground | Listed in Table 2 |
| $\alpha$ | Ratio of the adjacent grounded resistances | 1 |
| $q$ ($\hat{q}$) | (Dimensionless) Charges stored in a block | |
| $i_c$ ($\hat{i_c}$) | (Dimensionless) Current toward the capacitance | |
| $i_r$ ($\hat{i_r}$) | (Dimensionless) Current away from anode and pass through block resistance | |
| $I$ ($\hat{I}$) | (Dimensionless) Current from one block to ground | |
| $V_{SB}$ ($\widehat{V_{SB}}$) | (Dimensionless) Voltage of the whole spring-block system | |

Table 2. Information of the electrokinetic parameters of the six selected sets in the Chen-Ouillon-Sornette model.

| Set | $\hat{r}$ | $\hat{c}$ | $\hat{L}$ | Damping Region | $\zeta$ | $\eta$ | $\Delta$ | $\omega$ | $\tau_q$ | $Q_1$ (%) | $Q_2$ (%) | $Q_3$ (%) | $Q_4$ (%) | $b$ (slope) |
|---|---|---|---|---|---|---|---|---|---|---|---|---|---|---|
| A | 5 | 5 | 0.1 | OD1 | 10.04 | 2.40 | 91.20 |  | 4.08 | 6.85 | 93.15 | 26.32 | 73.68 | 2.12±0.39 |
| B | 5 | 5 | $\widehat{L_{c1}}$ | CD1 | 0.92 | 0.21 | 0 |  | 2.17 | 6.96 | 93.04 | 26.32 | 73.68 | 2.16±0.42 |
| C | 5 | 5 | 10 | UD | 0.14 | 0.024 | -0.0764 | 0.14 | 14.29 | 8.67 | 91.33 | 26.32 | 73.68 | 2.05±0.24 |
| D | 5 | 5 | 100 | UD | 0.05 | 0.0024 | -0.0071 | 0.04 | 40 | 15.95 | 84.05 | 21.49 | 78.51 | 1.77±0.04 |
| E | 5 | 5 | $\widehat{L_{c2}}$ | CD2 | 0.0418 | 4.37E-4 | 0 |  | 47.85 | 13.60 | 86.40 | 18.42 | 81.58 | 1.72±0.03 |
| F | 5 | 5 | 700 | OD2 | 0.0414 | 3.43E-4 | 3.45E-4 |  | 87.62 | 13.17 | 86.83 | 16.23 | 83.77 | 1.72±0.03 |

Note: (i) $\widehat{L_{c1}}$=~1.1387 and $\widehat{L_{c2}}$=~548.8613. (ii) $Q_1+Q_2$=100% and $Q_3+Q_4$=100%, respectively. (iii) The range of the $b$ values is mean ± 1 standard deviation.